\documentclass[prl,aps,twocolumn,floatfix,nofootinbib,showpacs,superscriptaddress]{revtex4-1}

\usepackage{amsmath,amsfonts,amssymb,slashed,bm,graphicx,color}

\usepackage{varioref,MnSymbol}

\begin{document}

\title{A Divergence-Free Method to Extract Observables from Meson  Correlation Functions}
\author{$\text{Si-xue Qin}$}
\email{sixueqin@th.physik.uni-frankfurt.de}
\affiliation{Institute for Theoretical Physics, Johann Wolfgang Goethe University,
Max-von-Laue-Str.\ 1, D-60438 Frankfurt am Main, Germany}

\begin{abstract}
Correlation functions provide information on the properties of mesons in vacuum and of hot nuclear matter. In this Letter, we present a new method to derive a well-defined spectral representation for correlation functions. Combining this method with the quark gap equation and the inhomogeneous Bethe-Salpeter equation in the rainbow-ladder approximation, we calculate in-vacuum masses of light mesons and the electrical conductivity of the quark-gluon plasma. The analysis can be extended to other observables of strong-interaction systems. 
\end{abstract}


\date{\today}

\maketitle

\noindent\textbf{Introduction}. Hadrons contribute to most of the visible matter in our real world and are undoubtedly an embodiment of dynamical chiral symmetry breaking (DCSB) and confinement. Current and future hadron physics facilities are focusing on hadron spectroscopy in order to shed light on the mysteries of quantum chromodynamics (QCD). On the other hand, it is believed that the Relativistic Heavy Ion Collider (RHIC) and the Large Hadron Collider (LHC) are able to create the quark-gluon plasma (QGP) state of the early Universe through a ``mini-big bang". This provides us with the possibility to study quark-gluon dynamics directly and to enrich our understanding of the QCD phase diagram. The transport coefficients of the QGP, which directly reflect details of the quark-gluon interaction, are highly interesting from both experimental and theoretical viewpoints.

A unified description for physics in the two areas has been a central goal and great challenge for decades. Lattice QCD which is based on Monte Carlo simulations of quantum fields on finite discrete spacetime lattices has achieved numerous significant results, however, it also has its own limitations \cite{Fodor:2012gf,Philipsen:2007rj}. Thus, relativistically covariant formalisms of continuum quantum field theory (QFT) are still desirable. Among them, the Dyson-Schwinger equation (DSE) approach \cite{Krein:1990sf,Roberts:1994dr,Roberts:2007ji} is a framework that includes both DCSB and confinement \cite{McLerran:2007qj}. Remarkably, at zero temperature, $T=0$, a single DSE interaction kernel preserving the one-loop renormalisation group behavior of QCD has been able to provide a unified description of the pion'ís electromagnetic form factor \cite{Chang:2013nia}, its valence-quark distribution amplitude \cite{Chang:2013pq}, and numerous other quantities \cite{Maris:2003vk,Chang:2011vu}. Therefore, it is of great significance to extend the DSE approach to further quantitative studies of hadron and QGP physics.

In the DSE framework, hadrons, i.e., color-singlet bound states of quarks, are described by the Bethe-Salpeter equation (BSE) or the Faddeev equation. Solving these equations requires the quark propagator, i.e., the solution of the gap equation, on the complex momentum plane. The analytical structure of the quark propagator strongly depends on the specified truncation scheme and interaction model. As we know, it is difficult to solve bound-state equations which involve singularities of the quark propagator. This means that, in practical calculations, there always exists an upper limit for the masses of bound states, beyond which an extrapolation is necessary. However, it cannot be guaranteed that the extrapolation is always stable and reliable. At nonzero temperature, $T\neq 0$, Matsubara frequencies are introduced in imaginary-time thermal field theory \cite{LeBellac:2000wh}. Then, the situation is even more complicated since we do not know how to analytically continue Matsubara frequencies. Thus, it is a long-standing challenge to study light hadrons in vacuum with masses $>1$ GeV, bound states of light and heavy quarks, and hadrons in medium.

At $T\ne0$, transport coefficients can be calculated from meson spectral functions through Kubo formulae. Solving for meson spectral functions, one has to calculate Euclidean meson correlation functions. However, in terms of Green functions, the calculations are highly divergent. As we will see, the subtraction scheme which works at $T=0$ is not applicable at $T\ne0$. Thus, the divergence problem precludes the study of transport properties.

Herein we propose a novel approach based on spectral analysis, which can systematically solve the problems mentioned before. Using our new approach, we can extend the DSE study to a much wider range of applications. To demonstrate this, we calculate the masses of the $\pi$- and $\rho$-meson in vacuum and the electrical conductivity of the QGP with a single DSE interaction kernel. Both the result for the electrical conductivity and the approach itself  are essentially new.

\noindent\textbf{Meson Correlation Functions}. The retarded correlation function of local meson operators is defined as
\begin{eqnarray}
\Pi_H^R(t,\vec{x})=\langle J_{H}(t,\vec{x})J_{H}^\dagger(0,\vec{0}) \rangle_{\beta},
\end{eqnarray}
where $\beta=1/T$ and $\langle...\rangle_\beta$ denotes a thermal average. The operator $J_H$ has the following form
\begin{eqnarray}
J_H(t,\vec{x})=\bar{q}(t,\vec{x})\gamma_H q(t,\vec{x}),
\end{eqnarray}
with $\gamma_H=\mathbf{1},\gamma_5,\gamma_\mu,\gamma_5\gamma_\mu$ for scalar, pseudo-scalar, vector, and axial-vector channels, respectively. The meson spectral function is related to the imaginary part of the Fourier transform of the retarded meson correlation function
\cite{Blaizot:2001nr},
namely,\begin{eqnarray}
\rho_H(\omega,\vec{p})=2\,{\rm Im}\,\Pi_H^R(\omega,\vec{p}).
\label{eq:img}
\end{eqnarray}
Note that the spectral function is positive semi-definite for positive frequency and that $\rho_H(\omega,\vec{0})=-\rho_H(-\omega,\vec{0})$. In the zero-momentum limit, $\vec{p}\,=\vec{0}\,$, the Euclidean correlation function which can be connected with the retarded correlation function by analytic continuation, i.e., $\omega+i\epsilon\to i\omega_n$, has the following spectral representation,
\begin{eqnarray}
\Pi_H(\omega_n^{2})=\int_{\omega^2}^{\infty} \frac{\rho_H(\omega)}{\omega^2 +\omega_n^2}-({\rm
subtraction}),
\label{eq:spec1}
\end{eqnarray}
where $\int_{\omega^2}^{\infty}=\int_{0}^{\infty}\frac{d\omega^2}{2\pi}$; $\omega_n=2n\pi T, n\in Z$, are the bosonic Matsubara frequencies. Note that an appropriate subtraction is required because the spectral integral in Eq.\ \eqref{eq:spec1} does not converge, i.e., $\rho_H(\omega\to\infty)\propto \omega^2$.

Using the Fourier transform on Eq. \eqref{eq:spec1}, one can obtain the spectral representation of the Euclidean temporal correlation functions  without any subtraction. Lattice QCD generally adopts such a form \cite{Petreczky:2003iz}. However, it is not applicable for the DSE approach. As we will see, the numerical calculation of the Fourier transform is actually very difficult because of divergences in computing $\Pi_H(\omega_n^{2})$ by the DSE approach. At $T=0$, one has the so-called twice-subtracted dispersion relation \cite{Donoghue:1996kw} which is well-defined. At $T\ne0$, its straightforward extension reads 
\begin{equation}
\begin{split}
\Pi_H(\omega_{n}^{2})=\Pi_H(0)+\omega_{n}^2\Pi'_H(0)+\int_{\omega^2}^{\infty}
\frac{\omega_{n}^4\rho_H(\omega)}{\omega^4(\omega^2
+\omega_{n}^2)}.\label{eq:twicesb}
\end{split}
\end{equation}
The above equation takes care of the ultraviolet divergence. However, it generates a divergence in the infrared region because $\rho_{H}(\omega\to0)\propto\omega$ at $T\ne0$. Moreover, Eq. \eqref{eq:twicesb} is correct only if the derivatives of the Euclidean and retarded correlators can be connected by analytical continuation. It can be proved that such an analytical continuation does not hold at $T\ne0$. At one-loop level, one can easily check that the analytical continuation breaks down for the zeroth component of the vector correlation function. Thus, Eq. \eqref{eq:twicesb} is ill-defined and useless.

Here we would like to present a new method to construct a well-defined spectral representation. We introduce a transform for a function $f(x)$,
\begin{eqnarray}
\hat{\mathcal{O}}_N (x_{1},...,x_{N})\{f\}=\sum_{i=1}^{N}f(x_{i}) \prod_{j\ne i}^{N}\frac{1}{x_i-x_{j}}\,,
\label{eq:transform}
\end{eqnarray}
where $x_1\ne x_2\ne...\ne x_N$. If $f(x)$ is an $N$-order polynomial, then $\hat{\mathcal{O}}_{N+2}\{f\}=0$, e.g., $\hat{\mathcal{O}}_3\{{\rm linear\ function}\}=0$. According to analytical properties of correlation functions in QFT \cite{Peskin:1995ev}, the subtractions in the dispersion relations are always polynomials of momenta (or Matsubara frequencies), e.g., the subtraction for the meson correlation function is a linear function of $\omega_n^2$. Thus, using the $3$-order transform for $\Pi_H$, i.e., $\hat\Pi_H(\omega_{i}^{2},\omega_{j}^2,\omega_{k}^2)=\hat{\mathcal{O}}_{3} (\omega_{i}^2,\omega_{j}^2,\omega_{k}^2)\{\Pi_H\}$, one can obtain the following spectral representation,
\begin{eqnarray}
\hat\Pi_H(\omega_{i}^{2},\omega_{j}^2,\omega_{k}^2)=\int_{\omega^2}^\infty \frac{\rho_{H}(\omega)}{(\omega^2+\omega_i^2)(\omega^2+\omega_j^2)(\omega^2+\omega_k^2)},\quad
\label{eq:spec}
\end{eqnarray}
which is consistent with Eq.\ \eqref{eq:img} by analytic continuation. Note that Eq.\ \eqref{eq:spec} is well-defined and has no such problems as Eq.\ \eqref{eq:twicesb}. For numerical convenience, one can introduce a one-variable correlator as $\tilde\Pi_H(\omega_{i}^{2})=\hat\Pi_H(\omega_{i}^2,\omega_{i+1}^2,\omega_{i+2}^2)$ or $\tilde\Pi_H(\omega_{i}^{2})=\hat\Pi_H(\omega_{i}^2,\omega_{\rm fixed}^2,\omega_{\rm fixed}^2)$, which  reduces Eq.\ \eqref{eq:spec} to a one-dimensional equation.

\noindent\textbf{Dyson-Schwinger Equations}. In terms of Green functions, the Euclidean meson correlation function, $\Pi_H(\omega_{n}^{2})$, is defined as
\begin{eqnarray}
\parbox{70mm}{\includegraphics[width=\linewidth]{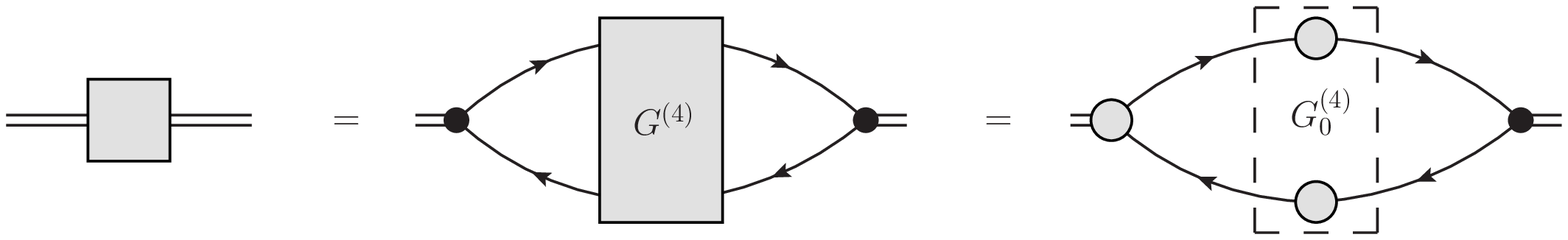}}\,,
\label{eq:mesoncor0}
\end{eqnarray}
where gray circular blobs denote dressed propagators and vertices; $G^{(4)}$ denotes the full quark--anti-quark four-point Green function; $G_0^{(4)}$ denotes the two disconnected dressed quark propagators in the dashed box; black dots denote bare propagators or vertices. One can easily check that the loop integral of Eq.\ \eqref{eq:mesoncor0} is highly divergent, which directly leads to the numerical difficulty in computing its Fourier transform. However, because Eq. \eqref{eq:spec} is an exact expression and its right-hand side is a well-defined integral, its left-hand side is automatically free from divergences. Thus, using the $3$-order transform for Eq.\ \eqref{eq:mesoncor0}, one can obtain a well-defined correlation function, $\hat\Pi_H$, which has a spectral representation as Eq.\ \eqref{eq:spec}.

The dressed propagators and vertices in Eq.\ \eqref{eq:mesoncor0} can be solved by the corresponding DSE. Using the abelian rainbow-ladder (RL) truncation, the  gap equation for the quark propagator is written as
\begin{eqnarray}
&& S(\tilde\omega_n,\vec{p}\,)^{-1} = Z_2(i\vec{\gamma}\cdot\vec{p} + i\gamma_4 \tilde\omega_{n} + Z_m m)\notag\\
  &&+\, Z_1\displaystyle\sumint_q {g^{2}} D_{\mu\nu} (\vec{k}, \Omega_{nl})\frac{\lambda_a}{2}{\gamma_{\mu}} S(\tilde\omega_l,\vec{q}\,)\frac{\lambda_a}{2}\gamma_\nu,\quad
\label{eq:gap}
\end{eqnarray}
where $\tilde\omega_l=(2l+1)\pi T, l\in Z$, are the fermionic Matsubara frequencies; $\sumint_q = T \sum_l\int\frac{d^3\vec{p}}{(2\pi)^3}$ denotes the Matsubara summation and the spatial momentum integral; $Z_{1,2,m}$ are the vertex, quark wave-function, and mass renormalization constants, respectively; $D_{\mu\nu} (\vec{k}, \Omega_{nl})$, with $(\vec{k},\Omega_{nl})=(\vec{p}-\vec{q},\tilde\omega_n-\tilde\omega_l)$, is the dressed gluon propagator. The inhomogeneous BSE for the dressed vertex is written as
\begin{eqnarray}
&& \Gamma_H(\omega_n;\tilde\omega_m,\vec{p}\,) = Z_H\gamma_H - Z_1\displaystyle\sumint_q {g^{2}} D_{\mu\nu} (\vec{k},\Omega_{ml})\notag\\
  &&\times\,\frac{\lambda_a}{2}{\gamma_{\mu}}S(\tilde\omega_l,\vec{q}\,)\Gamma_H(\omega_n;\tilde\omega_l,\vec{q}\,) S(\tilde\omega_l+\tilde\omega_n,\vec{q})\frac{\lambda_a}{2}\gamma_\nu,\quad
\label{eq:bse}
\end{eqnarray}
where the renormalization constant $Z_H$ is, respectively, $Z_4$ ($=Z_2Z_m$) and $Z_{2}$ for the (pseudo-)scalar and the (axial-)vector. Note that, since the RL truncation is  the leading term in a symmetry-preserving truncation scheme, the solutions of Eqs. \eqref{eq:gap} and \eqref{eq:bse} satisfy Ward-Takahashi identities \cite{Ball:1980ay,Qin:2013mta}.

The gap equation and the inhomogeneous BSE are fully determined by a specified interaction model, i.e., $g^2D_{\mu\nu} (\vec{k}, \Omega_{nl})$. Following Ref. \cite{Qin:2011dd}, we employ a one-loop renormalization-group-improved interaction model, which has two parameters: a strength $D$ and a width $\xi$. With the product $D\xi$ fixed, one can  obtain  a uniformly good description of pseudoscalar and vector mesons  in vacuum with masses $\lesssim1$ GeV if $\xi\in[0.4,0.6]$ GeV. We use $\xi=0.5$ GeV. In the QGP region, we follow Ref. \cite{Qin:2010pc} to include a Debye mass in the longitudinal projection of the gluon propagator and a logarithmic screening for the nonperturbative interaction. With the thermally modified model, one can obtain that the thermal quark masses for massless quarks are proportional to temperature for a very hot QGP, i.e., $m_T=0.8T$ for $T\gtrsim3T_c$, which is consistent with lattice QCD \cite{Karsch:2009tp}.

\noindent\textbf{Extraction of Observables}. At $T=0$, the Euclidean meson correlation functions can be written in $O(4)$ covariant form. Then, we have the corresponding spectral representation,
\begin{eqnarray}
\hat\Pi_H(s,s_1,s_2)=\int_{m^2}^\infty\frac{\rho_{H}(m)}{(m^{2}+s)(m^{2}+s_1)(m^{2}+s_2)},\quad
\label{eq:zeroallrho}
\end{eqnarray}
The information on mesons of channel $H$ can be extracted from the spectral function $\rho_H(m)$; viz., the ground state and radially excited states correspond to peaks of $\rho_H(m)$.  

According to Eq. \eqref{eq:zeroallrho}, the whole tower of states, from the ground state to all radially excited states, contributes, but the ground state dominates. To isolate each bound state, we can insert the inhomogeneous BSE into Eq. \eqref{eq:mesoncor0} and obtain
\begin{eqnarray}
\parbox{70mm}{\includegraphics[width=\linewidth]{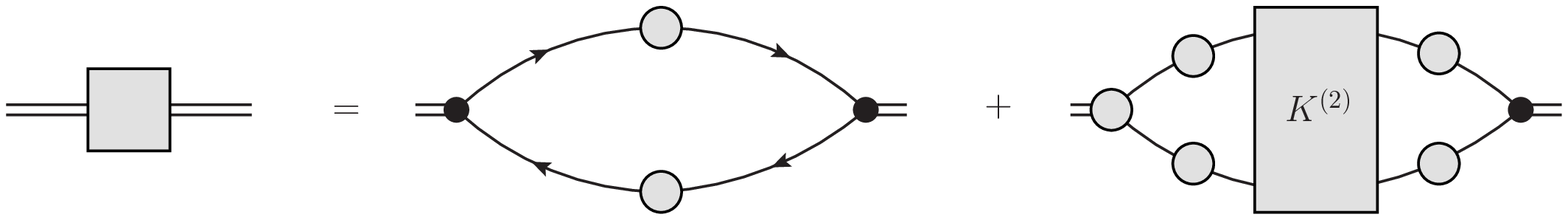}}\,.
\label{eq:eigende1}
\end{eqnarray}
The two-particle irreducible kernel $K^{(2)}$ has the following complete eigenvector decomposition
\begin{eqnarray}
G_0^{(4)} \cdot K^{(2)}\cdot G_0^{(4)}=\sum_n^{\infty}
\lambda_n |\chi_n\rangle\langle\chi_n| \,.
\label{eq:eigende2}
\end{eqnarray}
Note that the eigenvectors are normalized and orthogonal to each other. According to the homogeneous BSE, the eigenvector $|\chi_n\rangle$ is the wave function of a bound state if $\lambda_n=1$, and each eigenvalue only traces a single bound state. Thus, the meson correlation function can be projected onto the component of an individual bound state by inserting Eq. \eqref{eq:eigende2} into Eq. \eqref{eq:eigende1}.

At $T\ne0$, properties of in-medium mesons can be obtained by the corresponding spectral functions, and their dissociation can be read off  from the width of the corresponding peaks. Especially the vector spectral function which is related to the electromagnetic current correlation function is of significance for observables of the QGP, e.g., electrical conductivity, heavy-quark diffusion constant, thermal dilepton rate, etc.. From the Kubo formula, the electrical conductivity can be expressed as $\sigma_{\rm em}=\sigma e^2\sum_f Q_f$
(sum of the electrical charges of flavored quarks) and
\begin{eqnarray}
\sigma=\frac{1}{6}\lim_{\omega\to 0}\sum_{i=1}^{3}\frac{\rho_V^{ii}(\omega,\vec{p}\,=0)}{\omega},
\label{eq:elecsigma}
\end{eqnarray}
where $\rho_V^{ii}$ are the spatial components of the vector spectral function (in what follows, the summation is suppressed unless stated).

\begin{figure}
\centering
\includegraphics[width=0.9\linewidth]{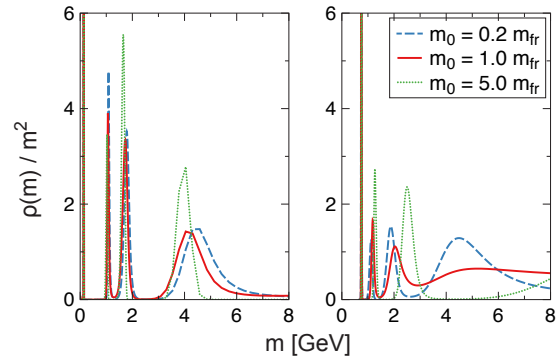}
\caption{(color online) Meson spectral functions of the pseudoscalar (left) and the vector (right) channels, and their sensitivities to the variation of the MEM default model.} \label{fig:spec0}
\end{figure}

\noindent\textbf{Numerical Results}. At $T=0$, we use the maximum entropy method (MEM) \cite{Bryan:1990tv,Nickel:2006mm,Mueller:2010ah} to solve for spectral functions from Eq. \eqref{eq:zeroallrho}. Following lattice QCD \cite{Asakawa:2000tr}, we choose the MEM default model as $m_0m^2$. For simplicity, we let $m_0=m_{\rm fr}$ which is calculated in the non-interacting limit \cite{Karsch:2003wy,Aarts:2005hg}. To check the sensitivity of the result to the default model, we vary $m_0$ by a factor $5$ as in Ref. \cite{Asakawa:2000tr}. The calculated pseudoscalar and vector spectral functions are plotted in Fig. \ref{fig:spec0}. It is found that the first peaks which correspond to the ground states of the $\pi$- and $\rho$-meson, are very sharp and robust against the variation of $m_0$ (uncertainties of their locations are $<1$ MeV). Compared with the result obtained by the homogeneous BSE, the ground-state masses are precise (see Table \ref{tab:zeromass}). 
\begin{table}
\centering
\caption{Masses of the $\pi$- and $\rho$-meson in vacuum and their comparison with the results obtained from the homogeneous BSE (h.BSE), where the upper errors are obtained with $m_0=0.2m_{\rm fr}$, while the lower errors are obtained with $m_0=5.0m_{\rm fr}$ [parameters I and II follow Refs. \cite{Qin:2011dd,Qin:2011xq}, and dimensional quantities are reported in GeV]. \label{tab:zeromass}}
\begin{tabular*}{1.0\linewidth}{@{\extracolsep{\fill}}cccccc}\hline\hline
method & para. & $\pi$  & $\rho$ & $\pi'$ & $\rho'$ \\\hline
this work & I &  0.135 &0.748 & $1.065^{+0.021}_{-0.025}$ &  $1.185^{-0.003}_{+0.045}$ \\
 h.BSE & I & 0.134 & 0.742 & 1.071 & 0.974 \\
 this work & II &  0.152\ & 1.043 & $1.461^{+0.024}_{-0.077}$  & $1.239^{-0.033}_{+0.015}$ \\
 h.BSE & II &  0.155 & 1.046 & 1.283 & 1.260 \\
 \hline\hline
\end{tabular*}
\end{table}

But the second and high-energy peaks are not so stable against the variation of the default model. For example, it is supposed that the second peaks correspond to the first radially excited states. However, compared with the result obtained by the homogeneous BSE, this is not the case all the time. We find that the MEM tends to merge close peaks. Sometimes, the second peaks are broad and do not truly correspond to the first radially excited states (see the last two columns in Table \ref{tab:zeromass}). This is an intrinsic drawback of the MEM and also happens in lattice QCD \cite{Asakawa:2000tr}, which can be improved by the high-precision MEM \cite{Rothkopf:2011ef}. Also, the eigenvector projection, i.e., Eq. \eqref{eq:eigende2}, can be adopted to solve the problem even more effectively.

\begin{figure}
\centering
\includegraphics[width=0.9\linewidth]{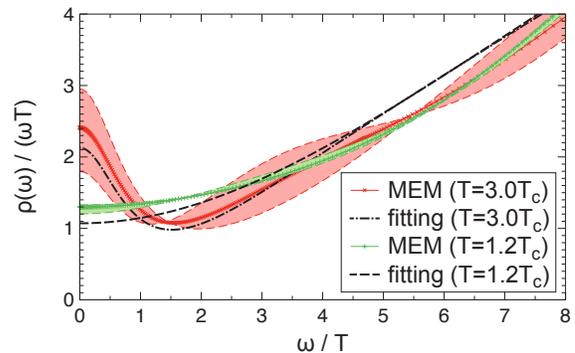}
\caption{(color online) Vector spectral functions at different temperatures (the solid lines are obtained by using the fitted spectral function as the default model directly, and the shaded ranges around the curves are obtained by altering the default model).} \label{fig:spect}
\end{figure}

At $T\ne0$ (and zero chemical potential), we implement calculations with the physical parameters I in Table \ref{tab:zeromass}. It is found that the hadron gas transits to the QGP at $T_c=144$ MeV \cite{Qin:2013ufa}.  At $T>T_c$, considering that light hadrons are dissolved in the QGP, we can parameterize the vector spectral function as,
\begin{eqnarray}
\rho_V^{ii}(\omega)=\frac{2\chi\omega}{\omega^2+\eta^2} + \frac{3}{2\pi}(1+\kappa)\omega^2 \tanh\left(\frac{\omega}{4T}\right),
\end{eqnarray}
where the first part is the Breit-Wigner-like (BW-like) distribution \cite{Aarts:2002cc,Ding:2010ga}, and the second part is the perturbative continuous tail \cite{Altherr:1989jc}. Note that there are three parameters in the above form. Inserting it into Eq. \eqref{eq:spec}, we can fit the parameters with a very high precision. The fitted vector spectral functions are plotted in Fig. \ref{fig:spect}. It is found that the BW-like distribution becomes higher and sharper with increasing $T$.

\begin{figure}
\centering
\includegraphics[width=0.9\linewidth]{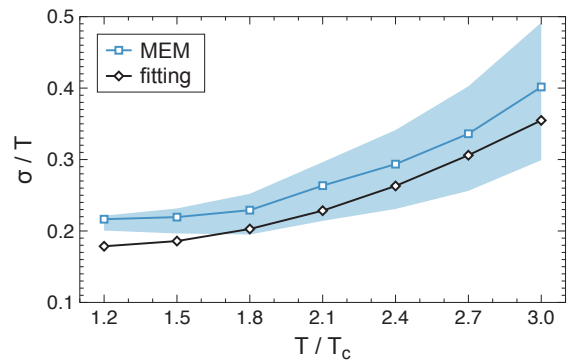}
\caption{(color online) Behavior of the electrical conductivity with temperature (the shaded range around the curve is obtained by altering the default model).} \label{fig:sigma}
\end{figure}

To check the reliability of the fitting, we use the fitted spectral function as the MEM default model, and analyze the sensitivity of the MEM output to the variation of the default model (by doubling or halving the BW-like part). The result is also plotted in Fig. \ref{fig:spect}. It is found that the spectral functions obtained by the MEM are close to the fitted ones and the uncertainties are tolerable. Finally, we calculate the electrical conductivity of the QGP and study its evolution with $T$. The result is plotted in Fig. \ref{fig:sigma}, which is consistent with the recent results of lattice QCD \cite{Ding:2010ga,Aarts:2007wj}. It is found that $\sigma/T$ increases with increasing $T$, which indicates that the coupling strength of the QGP decreases with increasing $T$.

\noindent\textbf{Epilogue}. We presented a novel approach which has no analyticity and divergence problems as other approaches, to systematically study meson spectroscopy in vacuum and transport properties of the QGP. Combining it with the solutions of the rainbow-ladder truncated DSE, we reproduced masses of the $\pi$- and $\rho$-meson ground states, and then calculated the electrical conductivity of the QGP for the first time. The magnitude of the electrical conductivity is comparable with recent results of lattice QCD, and its evolution with $T$ indicates a strongly-coupled QGP in the neighborhood of $T_c$. 

The key of the new approach is the transform  \eqref{eq:transform}, which can be easily extended to baryon correlation functions. According to Eq. \eqref{eq:eigende2}, the new approach is potentially applicable for radially excited states. Resonances which are not well defined in the homogeneous BSE can also be studied by the new approach. Thus, the new approach could provide the possibility to achieve a unified description of light hadrons with masses $<2$ GeV by the DSE approach. In medium, more transport coefficients, e.g., heavy quark diffusion constant, thermal dilepton rate, etc., can also be studied by the DSE approach. Moreover, the study is not limited to zero chemical potential. Therefore, the new approach can potentially connect observables with QCD phase transitions on the whole temperature--chemical-potential plane.

\noindent\textbf{Acknowledgements}.  The author would like to thank Y.-x.\ Liu, D.\ H.\ Rischke, and C.\ D.\ Roberts for helpful discussions. The work was supported by the Alexander von Humboldt Foundation through a Postdoctoral Research Fellowship.

\bibliography{MesCorFunQ}

\end{document}